\documentclass[preprint,cite,bm,figures]{epl}
\usepackage{graphicx}
\usepackage{amsmath}
\graphicspath{{.}{./EPS/}}

\title{Universal Rashba spin precession of two-dimensional\\
electrons and holes}
\shorttitle{Universal Rashba spin precession etc.}
\author{Marco G.~Pala\inst{1,2} \and Michele Governale\inst{2} \and 
J\"urgen K\"onig\inst{2} \and Ulrich Z\"ulicke \inst{2}}
\shortauthor{M. G. Pala \etal}
\institute{
\inst{1} Dipartimento di Ingegneria dell'Informazione, via
Diotisalvi 2, I-56126 Pisa, Italy\\
\inst{2} Institut f\"ur Theoretische Festk\"orperphysik,
Universit\"at Karlsruhe, D-76128 Karlsruhe, Germany
}


\date{\today}

\begin{document}

\maketitle

\begin{abstract}
We study spin precession due to Rashba spin splitting of electrons 
and holes in semiconductor quantum wells.
Based on a simple analytical expression that we derive for the current 
modulation in a broad class of experimental situations of
ferromagnet/nonmagnetic semiconductor/ferromagnet 
hybrid structures, we conclude
that the Datta-Das spin transistor (i) is feasible with holes and (ii)
its functionality is not affected by integration over injection angles.
The current modulation shows a universal oscillation period, irrespective 
of the different forms of the Rashba Hamiltonian for electrons and holes.
The analytic formulas approximate extremely well exact numerical 
calculations of a more elaborate Kohn--Luttinger model.
\end{abstract}

Transport effects based on coherent manipulation of the spin degree
of freedom in low--dimensional semiconductors are currently attracting 
a lot of attention~\cite{lossbook}.
These studies, enabled by recent progress in nanofabrication technology
to create high--quality samples, are motivated by both their 
interesting fundamental physics and their potential for future 
device applications~\cite{wolf}.
A lot of progress in the field has been stimulated by 
the exploitation of spin precession due to Rashba spin--orbit (SO) 
coupling in 2D systems 
both for electrons~\cite{rashba,byra:jpc:84,lommer}
and for holes~\cite{gerchikov}.
A prominent example is the spin--controlled field--effect transistor 
(spin FET) introduced by Datta and Das~\cite{spinfet}, followed by more recent 
proposals for novel devices utilizing Rashba SO coupling~\cite{devices}.
Both in the original~\cite{spinfet} and most subsequent~\cite{mireles,hausler,
matsuyama,rashbawire,egues,pareek} works, a quasi--onedimensional (1D)
confinement was considered essential for proper spin--FET action.
Spin precession in truly 2D electron systems was studied
numerically in a number of works~\cite{bruno, raichev}.
On the other hand,
ever-present spin relaxation will reduce the spin polarization of
currents, preventing the realization of gate-controlled modulation.
Such processes arise, e.g., from
magnetic impurities but most importantly from elastic impurity
scattering that randomizes the direction of the effective Rashba field.
Stronger spin-orbit coupling and band mixing phenomena
imply a shorter spin relaxation time for the holes respect to the electrons,
that can be compensated by shorter precession length.
A nice proposal, which exploits tunability of Rashba SO coupling, 
to overcome the detrimental effects due to scattering processes is
presented in the last paper of Ref.~\cite{devices}.

\begin{figure}
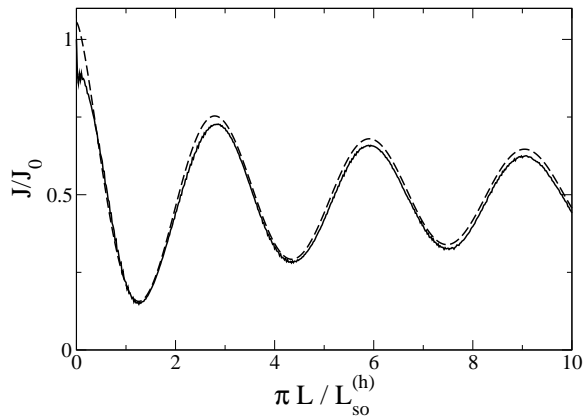

\onefigure[width=3in]{fig1.eps}
\caption{Current modulation in a p--type 2D
ferromagnet/nonmagnetic-semiconductor/ferromagnet hybrid system computed
numerically (solid line) for a realistic system and compared
with our analytic expression (dashed line). Here $L$ is the
width of the nonmagnetic region, and
$L_{\mathrm{so}}^{\mathrm{(h)}}$ the spin--precession length for
holes due to the Rashba effect. The magnetizations of the two ferromagnetic
regions are parallel and point in a direction perpendicular to the 2D plane 
of the hybrid system. For the numerical simulation we have used 
parameter values for a realistic \chem{GaMnAs}/\chem{GaAs}/\chem{GaMnAs} 
2D heterostructure. 
The hole densities are of order $10^{16} \text{ m}^{-2}$, 
$L_{\mathrm{so}}^{\mathrm{(h)}}
\approx 300 \text{ nm}$, and the exchange field is $150 \text{ meV}$. 
There is a Fermi wave vector mismatch 
$k_{\text{F}}^{(\text{f})}/k_{\text{F},0}=1.33$; and the Fermi wave vector   
splitting due to SO coupling is $\Delta k/k_{\text{F},0}=7.66\cdot10^{-3}$,
corresponding to 
$\beta_{\text{p}} \langle E_z \rangle= 1.3 \cdot 10^{20}$ 
eV$^{-1}$ sec$^{-2}$ nm.} 
\label{compplot}
\end{figure}

The aim of this Letter is twofold. First, we provide a unified
analytical description of spin precession for both electrons 
{\em and\/} holes in the case of ballistic transport regime, 
discussing common universal features and retaining
the 2D nature of the problem. This is the most realistic case
for anticipated device applications. One of our main results is
illustrated in Fig.~\ref{compplot} where our approximate
analytical expression for the current modulation in a 2D hole
spin FET is compared with the full numerical result obtained by
mode-matching within  a more elaborate Kohn--Luttinger model.
Except for very small values of the distance $L$ between the
ferromagnetic contacts, the agreement is excellent. We prove
analytically that the oscillation period in the 2D setup is the
same as in the quasi--1D structures considered
before~\cite{spinfet, mireles}. Second, we want to emphasize the
utility of holes as carriers in a spin FET. The apparent
disadvantage of a shorter spin life time as compared to electrons
is off--set by shorter spin--precession lengths and, above all, by
the possibility to use recently created~\cite{ohno} p--type
ferromagnetic semiconductors as current injectors. Besides obvious
advantages concerning integrability in current semiconductor
technology, such an {\em all--semiconductor spin FET} would circumvent the
problems~\cite{schmidt} that prevent spin injection from metallic
magnets into semiconductors in the absence of a large interface
barrier.

\begin{figure}
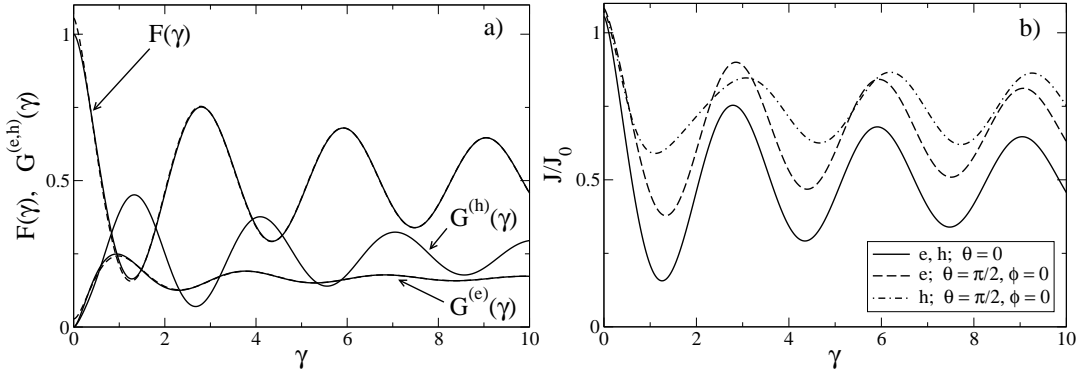

\twoimages[width=2.8in]{fig2a.eps}{fig2b.eps}
\caption{
a) Plot of the functions $F(\gamma)$ and
$G^{(\text{e,h})}(\gamma)$ computed numerically (solid lines) and
by means of the approximated formulae given in the text (dashed
line). The difference between analytical and numerical results is
only barely visible for $\gamma \to 0$. b) Current density  as a
function of $\gamma=\pi L/L_{\text{so}}^{(\text{e,h})}$ for
different magnetization directions for both the electron and hole
cases. The current is computed using the approximated expressions
for $F(\gamma)$ and $G^{(\text{e,h})}(\gamma)$.} 
\label{fgplots}
\end{figure} 
We now turn to explaining details of our calculations. A 2D hybrid
system with two semi--infinite ferromagnetic contacts separated by
a nonmagnetic 2D stripe of width $L$ is considered. To be
specific, the extension of both ferromagnetic and nonmagnetic
parts is infinite in $y$ direction, and the nonmagnetic stripe is
bounded by interfaces to the ferromagnetic contacts at $x=0$ and
$x=L$. Due to structural inversion asymmetry in the growth ($z$)
direction, charge carriers are subject to a spin--orbit coupling
of the Rashba type~\cite{rashba,lommer,gerchikov}. 
For electrons in the s--like conduction band, 
it reads $\beta_{\text{s}} \langle E_z
\rangle  (\vec{p}\times \vec{\sigma}) \cdot \hat{z}$, while for
the p--like valence bands we have $\beta_{\text{p}} \langle E_z
\rangle (\vec{p}\times \vec{J})\cdot \hat{z}$~\cite{winkler1,
winkler2}. Here, $\langle E_z \rangle$ is related to the average
of the electric field in growth direction.
In 2D hole quantum wells, the degeneracy of heavy--hole
(HH) and light--hole (LH) valence bands at the zone center is
lifted by the vertical confinement~\cite{chow}, and only the first
HH subband (HH$_1$) is populated for typical 2D sheet densities.
In the following, we consider only the HH$_1$ subband unless
specified otherwise. The effective Hamiltonians for Rashba SO
coupling experienced by electrons and holes are~\cite{winkler1,winkler2}
\begin{subequations}
\label{hso}
\begin{eqnarray}
\label{hse}
H_{\text{so}}^{(\text{e})}&=&\beta_{\text{s}} \langle E_z \rangle
i \left[p_{-}\sigma_{+}-p_{+}\sigma_{-}\right]
\\ 
\label{hsh}
H_{\text{so}}^{(\text{h})}&=&\beta_{\text{h}} \langle E_z \rangle
i \left[p_{-}^{3}\sigma_{+}-p_{+}^{3}\sigma_{-}\right],
\end{eqnarray}
\end{subequations}
where $p_{\pm}=p_x\pm i p_y$ are linear combinations of momentum
components, $\sigma_\pm=(\sigma_x\pm i \sigma_y)/2$ denote Pauli
matrices, and $\beta_{\text{s,h}}$ are material parameters. 

We assume identical ferromagnetic contacts having parallel
magnetization $M$ whose direction (in 3D) is described by two
polar angles $(\theta, \phi)$,
where $\phi=0$ corresponds to magnetization perpendicular to interface. 
As quasi-2D ferromagnets have either
easy--axis or easy--plane anisotropy, the two relevant cases are
that with magnetization perpendicular to the 2D plane
($\theta=0$) or in the plane ($\theta=\pi/2$). Assuming that there
is no reflection at the interfaces between the nonmagnetic semiconductor 
and the ferromagnetic contacts, we can calculate the transmission
probability $T_{\theta,\phi}(\alpha)$ for an electron impinging on
the first interface at an angle $\alpha$. 
(Details and a discussion of the approximation are given further below.)
The total current density is then $J\propto
\int_{-\pi/2}^{\pi/2} T_{\theta,\phi}(\alpha)\cos(\alpha)\,
d\alpha$. We find its expression for arbitrary 
$(\theta, \phi)$ as
\begin{equation}
J=J_0 \Big\{ \cos^2 \theta F(\gamma)+\sin^2 \theta   
\Big[
\sin^2 \phi+ F(\gamma)\cos^2 \phi+ 
 G^{(\text{e,h})}(\gamma)\cos (2 \phi)
\Big]\Big\},
\label{current}
\end{equation}
where the functions $F$ and  $G^{(\text{e,h})}$ are shown in the 
panel a) of Fig.~\ref{fgplots}, and are defined explicitly
below. Here $\gamma=\pi L /L_{\text{so}}^{(\text{e,h})}$, where
$L_{\text{so}}^{(\text{e,h})}=2 \pi/\Delta k$ is the spin precession 
length for electrons or holes, with $\Delta k$ being the difference 
of Fermi wave vectors for the spin--split 2D subbands:
\begin{subequations}
\label{lso}
\begin{eqnarray}
\label{lsoe}
L_{\text{so}}^{(\text{e})}&=& \frac{\pi} {\beta_{\text{s}} 
\langle E_z \rangle} 
\left(\frac{\hbar}{m^{(\text{e})}}\right)\quad ,   \\
\label{lsoh}
L_{\text{so}}^{(\text{h})}&=&\frac{\pi}{2\beta_{\text{h}}\langle E_z \rangle} 
\left(\frac{\hbar}{m^{(\text{h})}}\right)^2 \frac{1}
{\epsilon_{\text{F}}}\quad .
\end{eqnarray}
\end{subequations}
Here $m^{(\text{e,h})}$ is the effective mass for electrons/holes,
and $\epsilon_{\text{F}}$ is the Fermi energy.
Equation~(\ref{lsoh}) is valid when the Rashba spin splitting is small
compared to the Fermi energy. 
Equation~(\ref{current}) is one of the central results of this
Letter, and a few remarks on it are in order. Once we decide
whether the carriers are electrons or holes and, hence, fix 
the symmetry of the band (s or p type), the current density has a
universal behavior: it depends only on the ratio of the distance
between the ferromagnetic contacts and the spin-precession length.
Finite transparency of the interfaces and Fermi--velocity mismatch
between the ferromagnetic and nonmagnetic materials will be shown
below to lead only to small quantitative changes. The only difference
in the functional form of the current for electrons and holes is
in $G^{(\text{e,h})}(\gamma)$ and thus appears only if the
magnetization has an in-plane component. The current density
Eq.~(\ref{current}) shows oscillations in $\gamma$ with a period
$\pi$, i.e., oscillations in $L$ with a period
$L_{\text{so}}^{(\text{e,h})}$, that are a manifestation of spin
precession in the nonmagnetic region. A few observations can be
made regarding the behavior of the oscillatory part of the current
density for different magnetization directions (see panel b) of
Fig.~\ref{fgplots}): i)~The largest oscillations 
of the current (solid line in the plot) as a function of $\gamma$ 
occur for magnetization perpendicular to the 2D system ($\theta=0$);
ii)~In this case electrons and holes behave exactly in the same way; 
iii)~For in--plane magnetization 
perpendicular to the interface ($\theta=\pi/2$, $\phi=0$), 
the electron system (dashed line in the plot)
exhibits a slightly larger modulation 
of the conductance as compared to the hole system 
(dash-dotted line in the plot); 
iv) For in--plane magnetization parallel to the interface 
($\theta=\pi/2$, $\phi=\pi/2$) the conductance still shows
oscillations as a function of $\gamma$ (this is in contrast to the
case of normal incidence), which are larger for holes than electrons.
The different functional form for the SO term for electron and
holes [see Eq.(\ref{hso})] 
does not affect the transmission probability for
normal injection. Hence, the non universal features, when present, are
due to the integration over the injection angle $\alpha$.
We can conclude that the oscillations of the conductance as a function of
$L_{\text{so}}^{(\text{e,h})}$ are not washed out by the 2D geometry
of the system, i.e., by the necessity to integrate over the
direction of the incoming electrons in the ferromagnet. This
oscillatory behavior of the current is exploited by the spin FET.
Hence, our results show analytically that 
in the case of ballistic transport regime 
the spin FET {\em does not}
require a quasi--1D setup to work properly, and that the largest 
modulations of the conductance as a function of the spin
precession length are obtained for magnetization perpendicular to
the 2D system. Furthermore, Eq.~(\ref{current}) proves the
feasibility of a p-type spin--FET, with holes performing even
better than electrons for certain magnetization directions. 

To  illustrate the strategy employed to obtain  the general formula 
Eq.~(\ref{current}) without inflicting lengthy algebraic manipulations to 
the reader, it is sufficient to show the calculation for the case 
of a hole system and magnetization perpendicular to the plane ($\theta=0$).   
We start by calculating the probability for a plane wave in the 
left ferromagnet, impinging on the first interface at an angle
$\alpha$, to be transmitted in the second ferromagnet. We assume
that only majority spins contribute to transport, hence the 
spin state of the incoming electron in the left ferromagnet is
$|+\rangle$, where  $|+\rangle$ is the spinor corresponding to
spin in the $+z$ direction (the magnetization direction). We now
need to write the wave function in the semiconductor. We notice
first that the SO coupling Hamiltonian Eq.~(\ref{hsh}) removes 
the degeneracy of HH$_1$ subbands. The eigenstates are still plane
waves $\exp\left[i k (x \cos\alpha  +y \sin\alpha) \right]
\chi_{\pm}$, with spinors $ \chi_{\pm}= (1/\sqrt{2}) [|+\rangle
\mp i\exp(i 3\alpha)|-\rangle]$. The spin--split dispersion
relations associated to these states read $\epsilon_{\pm}
=\frac{\hbar^2}{2m^{(\text{h})}} k^2 \pm \beta_{\text{h}} 
\langle E_z \rangle \hbar^3 k^3$. As we study linear transport, we are
interested in the states at the Fermi energy. In particular, the
Fermi wave vectors for the $\epsilon_\pm$ bands are 
$k_{\text{F},\pm}=k_{\text{F},0}\mp \Delta k/2$ where 
$k_{\text{F},0}$ is the Fermi wave vector when no SO coupling is
present. Due to translational invariance along the interface, the
wave vector component parallel to the interface is conserved when
going from the ferromagnet to the non magnetic semiconductor. 
A plane wave in the 
ferromagnet with $\vec{k}= k_{\text F}^{(\text{f})}
(\hat{x} \cos\alpha + \hat{y} \sin \alpha)$ gives rise to two
transmitted waves with propagation directions defined by angles
$\alpha_{\pm}$. This effect is similar to birefringence~\cite{matsuyama,biref}:
the bands $\epsilon_\pm$ have different Fermi wave vectors, resulting in
two different propagating directions for the transmitted waves.
In the limit of weak SO coupling, i.e., $\Delta k / k_{\text{F},0}
\ll 1$, the two angles read $\alpha_{\pm}=\alpha_0\pm (1/2)(\Delta
k/k_{\text{F},0})\tan\alpha_0$, and $\alpha_0$ is defined by
$k_{\text{F}}^{(\text{f})}\sin\alpha=k_{\text{F},0}\sin\alpha_0$. 
We are now in the position to write the transmitted state in the 
nonmagnetic strip: it simply reads $c_{+} \exp \left[i  
k_{\text{F},+} (x \cos\alpha_+  +y \sin\alpha_+) \right] \chi_{+}
+c_{-} \exp \left[i k_{\text{F},-} (x \cos\alpha_{-}  +y \sin
\alpha_{-}) \right] \chi_{-}$. 
By assuming perfect transparency of the interface we can compute 
the coefficients $c_{\pm}$ simply by matching the wave functions
in the ferromagnet and non magnetic semiconductor. 
At the other interface $x=L$,
only the $|+\rangle$ component will be transmitted, hence the
outgoing state in the right ferromagnet reads $\exp \left[i 
k^{(\text{f})}_{\text{F}} (x \cos\alpha  +y \sin\alpha) \right] 
\cos[\Delta k L/ (2\cos \alpha_0)] |+\rangle$. From that the
transmission probability can be read off:
\begin{equation}
\label{tperp}
T_{0,\phi}(\alpha)=\cos^2\left[\frac{\gamma}{\cos \alpha_0}
\right]\quad , 
\end{equation} 
where we have used the relation $\Delta k\, L/2=\gamma$, and the
dependence on $\alpha$ is through $\alpha_0$ via the relation
$k_{\text{F}}^{(\text{f})}\sin\alpha=k_{\text{F},0}\sin\alpha_0$.
In a similar way we can obtain the transmission probabilities for
the electron case and for arbitrary magnetization direction. We
find that Eq.~(\ref{tperp}) is valid for both electrons and holes.
The transmission probability for in--plane magnetization reads 
\begin{equation}
\label{tinplane}
T_{\pi/2,\phi}(\alpha)=\cos^2\left[\frac{\gamma}{\cos \alpha_0}
\right] + \sin^2\left[\nu_{\text{e,h}}\alpha_0-\phi\right]\sin^2
\left[\frac{\gamma}{\cos \alpha_0}\right],  
\end{equation} 
where $\nu_{\text{e}}=1$ and $\nu_{\text{h}}=3$. 
Finally, we can write the transmission for arbitrary magnetization 
direction as
\begin{equation}
\label{tgeneral}
T_{\theta,\phi}(\alpha)=\cos^2\theta\, T_{0,\phi}+\sin^2\theta\,
T_{\pi/2,\phi}.
\end{equation} 
Equations~(\ref{tperp}--\ref{tgeneral}) cease to be valid once
one of the transmitted states in the nonmagnetic strip becomes evanescent
and is totally reflected. This condition defines 
the critical angles $\alpha_{\text{c},\pm}$, that in the limit of weak
SO coupling read  $\alpha_{\text{c},\pm}=\frac{k_{\text{F},0}}
{k_{\text{F}}^{(\text{f})}}\mp \frac{1}{2}\frac{\Delta k}
{k_{\text{F}}^{(\text{f})}}\approx\frac{k_{\text{F},0}}
{k_{\text{F}}^{(\text{f})}}=\alpha_{\text{c}}$.
At this point a few remarks on the approximate analytical treatment are 
in order: i) In this calculation scheme the spin precession in the 
nonmagnetic strip and the spin selecting properties of the ferromagnets 
(they act  as polarizer and analyzer)
are fully taken into account, while the interference effects arising from
multiple reflection between the interfaces are  neglected. 
This is in the same spirit of the calculation 
of Datta and Das \cite{spinfet} and of all the quantum-mechanical 
\textit{Gedanken} experiments involving polarized photons and polarizers. 
Indeed, multiple reflection introduces a modulation of the 
transmission coefficients $\propto \cos^2(L k_{\text{F}}/\cos\alpha)$. 
ii) This fast oscillation is washed out by the integration over the injection 
angle. This explains the remarkable agreement with the full 
quantum-mechanical calculation, shown in Fig.~\ref{compplot}.

The current density perpendicular to the interface is proportional
to $\int_{-\alpha_{\text{c}}}^{\alpha_{\text{c}}} T_{\theta,\phi}
(\alpha)\cos\alpha\, d\alpha\propto\int_{-\pi/2}^{\pi/2}
\tilde{T}_{\theta,\phi}(\alpha_0)\cos\alpha_0\,d\alpha_0$, where
$\tilde{T}_{\theta,\phi}(\alpha_0)=T_{\theta,\phi}(\alpha(\alpha_0
))$. Writing this integral with $T_{\theta,\phi}(\alpha)$ given in
Eqs.~(\ref{tperp}--\ref{tgeneral}), we obtain Eq.~(\ref{current}),
where $F(\gamma)$ and $G^{(\text{e,h})}$ are 
\begin{eqnarray}
\label{fint}
&F(\gamma)=\frac{1}{2}\int_{-\frac{\pi}{2}}^{\frac{\pi}{2}} \cos{\alpha} 
\cos^2{\left(\frac{\gamma}{\cos{\alpha}}\right)}\, d\alpha\, , &\\
\label{gint}
&G^{(\text{e,h})}(\gamma)= 
\frac{1}{2}\int_{-\frac{\pi}{2}}^{\frac{\pi}{2}} \cos{\alpha} 
\sin^2{(\nu_{\text{e,h}}\alpha)}
\sin^2{\left(\frac{\gamma}{\cos{\alpha}}\right)}\,d\alpha.&   
\end{eqnarray}
We have obtained an approximate analytical solution of
Eqs.~(\ref{fint}) and (\ref{gint}) that is valid unless $\gamma$
is very small:
\begin{eqnarray}
\nonumber
F(\gamma) &= &\frac{\pi \sqrt{2}}{8} \left\{ (4 \gamma-1) 
\text{S}\left( 2 \sqrt{\frac{\gamma}{\pi}}\right)
 +(-4 \gamma +1) \text{C}\left( 
2 \sqrt{\frac{\gamma}{\pi}}\right)+\right.\\
 \label{fapr}
 & &\left. 2 \sqrt{\frac{\gamma}{\pi}} \left[
\cos (2 \gamma)+\sin (2 \gamma)\right] + 1 +\frac{4}{\pi \sqrt{2}}
\right\}\quad ,\\
\nonumber
G^{(\text{e,h})}(\gamma)&=&\sqrt{2} \pi \left\{ P^{(\text{e,h})}(\gamma)
 \text{C}\left(
2 \sqrt{\frac{\gamma}{\pi}}\right)
+P^{(\text{e,h})}(-\gamma) \text{S}\left(
2 \sqrt{\frac{\gamma}{\pi}}\right)-\right.\\
\label{gapr}
& & \left.
\sqrt{\frac{\gamma}{\pi}}\left[Q^{(\text{e,h})}(\gamma) \cos(2\gamma)+
Q^{(\text{e,h})}(-\gamma) \sin(2\gamma)\right]+
M^{(\text{e,h})}(\gamma)\right\},
\end{eqnarray}
where $C$ and $S$ are cosine and sine Fresnel integrals~\cite{abramowitz}, 
respectively, and  
\begin{subequations}
\begin{eqnarray}
&  &P^{(\text{e})}(\gamma)=
\frac{2}{3} \gamma^3-\frac{1}{2} \gamma^2
+\frac{1}{8} \gamma + \frac{1}{32}\, ,\\  
& & P^{(\text{h})}(\gamma)=\frac{61}{315} \gamma^7 -
\frac{16}{45} \gamma^6 +\frac{4}{3} \gamma^5 -
\gamma^4 +\frac{11}{12} \gamma^3 -\frac{7}{16} \gamma^2 +
\frac{9}{64} \gamma + \frac{13}{256}\, ,\\
& &Q^{(\text{e})}=\frac{1}{3} \gamma^2
+ \frac{1}{6} \gamma + \frac{1}{16}\, ,\\ 
& &Q^{(\text{h})}=\frac{32}{315} \gamma^6 +
\frac{16}{105} \gamma^5 +
\frac{218}{315} \gamma^4 +
\frac{34}{105} \gamma^3 +
\frac{11}{24} \gamma^2 +
\frac{7}{48} \gamma + \frac{13}{128}\, , \\
& &M^{(\text{e})}(\gamma)= \frac{1}{2} \gamma^2
- \frac{1}{32} + \frac{1}{6 \pi \sqrt{2}}\, ,\\ 
& &M^{(\text{h})}(\gamma)=
\frac{16}{45} \gamma^6 + \gamma^4 +
\frac{7}{16} \gamma^2 + \frac{17 \sqrt{2}}{140 \pi} -
\frac{13}{256}\, .
\end{eqnarray}
\end{subequations}
In Fig.~\ref{fgplots} we compare the approximate results 
Eqs.~(\ref{fapr}) and (\ref{gapr}) with the integrals of 
Eqs.~(\ref{fint}) and (\ref{gint}). The
approximate formulae work extremely well: the largest errors are
for $\gamma=0$, and hence not in the region of physical interest,
and the errors decrease monotonically with increasing $\gamma$. 
The \textit{relative} errors for $\gamma=\pi$, i.e., after a full
precession, are of order $10^{-3}$~\cite{note1}. 

We have derived the above results for transmission probabilities 
[Eqs.~(\ref{tgeneral})] and the current density
[Eq.~(\ref{current})] making use of several approximations,
such as  perfectly transparent interfaces and,
in the hole case, the truncation to the lowest heavy-hole subband.
We now test the accuracy of our approximate
results by comparing them to a more realistic model for the
hole system which is the more complicated one. Both the 
magnetic and nonmagnetic parts of the hybrid system are modeled
by a $4\times 4$ Kohn-Luttinger Hamiltonian with HH and LH bands. 
The confinement in the growth direction is treated by means of the
envelope--function approximation. The ferromagnetic contacts are
described within the Stoner approach by an exchange field. 
Calculation of transmission and reflection coefficients by means
of a full quantum--mechanical mode matching at the two interfaces
and integration over angle of incidence yields the current density
shown in Fig.~\ref{compplot} as a solid line. Excellent agreement
between our approximate analytical results and the exact numerical curve is
apparent (note that for the latter a Fermi wave vector mismatch 
across the interface is present.) The only effect of  nonideal interfaces
on the current modulation is a renormalization of its amplitude $J_0$.

In conclusions, we have studied spin precession in electron and
hole system, finding a universal expression for the current in a 
spin FET geometry. The full 2D nature of the problem is retained. 
Our results show that the current through the system is determined
only by the common magnetization direction in the ferromagnets and
by the ratio of the distance between the ferromagnetic contacts
and the spin--precession length.    
\acknowledgments
This work was supported by DFG through the Center for Functional 
Nanostructures. M.~P.\ acknowledges support from the IST NanoTCAD
project (EC contract IST-1999-10828). We thank P. Mastrolia for
suggesting a method to evaluate the integrals in Eqs.~(\ref{fint})
and (\ref{gint}).


\begin{thebibliography}{25}
\bibitem{note1} 
The absolute errors at $\gamma=0$ for  $F(\gamma)$,
$G^{(\text{e})}(\gamma)$, and $G^{(\text{h})}(\gamma)$ are,
respectively,  $\pi/(4 \sqrt{2})-1/2\approx 5.54\cdot10^{-2}$, 
$1/6-\pi/(16\sqrt{2})\approx 2.78\cdot 10^{-2}$, and 
$13/70-13 \pi/(128 \sqrt{2})\approx 1.72\cdot 10^{-2}$. 
For $\gamma=\pi$, the relative errors for $F(\gamma)$,
$G^{(\text{e})}(\gamma)$, and $G^{(\text{h})}(\gamma)$ 
are, respectively, $1.83\cdot 10^{-3}$, $7.49\cdot 10^{-3}$, and 
$7.19\cdot 10^{-4}$.
\end{thebibliography}
\end{document}